\documentclass{ws-procs9x6}
\usepackage{amsmath}
\usepackage{epsfig}
\begin{document}
\title{HOW TO MEASURE THE CHARM DENSITY IN THE PROTON AT EIC
\footnote{Talk given at the Workshop on Exclusive Reactions at 
High Momentum Transfer (IV), May 18-21, 2010, TJNAF, Newport News, VA and Electron-Ion Collider Collaboration Meeting, July 29-31, 2010, Catholic University of America, Washington, DC}}
\author{N.Ya.~IVANOV}
\address{Yerevan Physics Institute, Alikhanian Bros. Str.~2, 0036 Yerevan, Armenia\\
E-mail: nikiv@yerphi.am\\
www.yerphi.am}


\begin{abstract}
\noindent We study two experimental ways to measure the heavy-quark content of the proton: 
using the Callan-Gross ratio $R(x,Q^2)=F_L/F_T$ and/or the azimuthal $\cos(2\varphi)$ asymmetry 
in deep inelastic lepton-nucleon scattering. Our approach is based on the 
following observations. First, unlike the production cross sections, the ratio $R(x,Q^2)=F_L/F_T$ and the 
azimuthal $\cos(2\varphi)$ asymmetry in heavy-quark leptoproduction are sufficiently stable, both 
parametricallly and perturbatively, in a wide region of variables $x$ and $Q^2$ within the 
fixed-flavor-number scheme of QCD. Second, both these quantities, $R(x,Q^2)=F_L/F_T$ and $\cos(2\varphi)$ 
asymmetry, are sensitive to resummation of the mass logarithms of the type $\alpha_{s}\ln\left( Q^{2}/m^{2}\right)$ 
within the variable-flavor-number schemes. These two facts together imply that the heavy-quark densities 
in the nucleon can, in principle, be determined from high-$Q^2$ data on the Callan-Gross ratio 
and/or the azimuthal asymmetry in heavy-quark leptoproduction. In particular, the charm content of the proton 
can be measured in future studies at the proposed Large Hadron-Electron (LHeC) and Electron-Ion (EIC) Colliders.

\end{abstract}
\keywords{Perturbative QCD, Heavy-Quark Leptoproduction, Mass Logarithms Resummation, 
Callan-Gross Ratio, Azimuthal Asymmetry}
\bodymatter

\section{Introduction}

The notion of the intrinsic charm (IC) content of the proton has been introduced about 30 years ago
in Ref.~[\refcite{BHPS}]. It was shown that, in the light-cone Fock space picture
\cite{brod1}, it is natural to expect a five-quark state contribution, $\left\vert
uudc\bar{c}\right\rangle$, to the proton wave function. This component can be generated by
$gg\rightarrow c\bar{c}$ fluctuations inside the proton where the gluons are coupled to different
valence quarks. The original concept of the charm density in the proton \cite{BHPS} has
nonperturbative nature since a five-quark contribution $\left\vert uudc\bar{c}\right\rangle$ scales
as $1/m^{2}$ where $m$ is the $c$-quark mass \cite{polyakov}.

In the middle of nineties, another point of view on the charm content of the proton has been proposed 
in the framework of the variable-flavor-number scheme (VFNS) \cite{ACOT,collins}.  The VFNS is an 
approach alternative to the traditional fixed-flavor-number scheme (FFNS) where only light degrees of
freedom ($u,d,s$ and $g$) are considered as active. Within the VFNS, the
mass logarithms of the type $\alpha_{s}\ln\left( Q^{2}/m^{2}\right)$ are resummed through the all
orders into a heavy quark density which evolves with $Q^{2}$ according to the standard DGLAP \cite{grib-lip}
evolution equation. Hence this approach introduces the parton distribution functions (PDFs) for the
heavy quarks and changes the number of active flavors by one unit when a heavy quark threshold is
crossed. Note also that the charm density arises within the VFNS perturbatively via the
$g\rightarrow c\bar{c}$ evolution. Some recent developments concerning the VFNS are presented in
Refs.~[\refcite{chi,SACOT,Thorne-NNLO}]. So, the VFNS was introduced to resum the mass logarithms 
and to improve thus the convergence of original pQCD series.

Presently, both nonperturbative IC and perturbative charm density are widely used for a
phenomenological description of available data. (A recent review of the theory and experimental
constraints on the charm quark distribution may be found in Ref.~[\refcite{pumplin}]). In particular, 
practically all the recent versions of the CTEQ \cite{CTEQ4,CTEQ5,CTEQ6} and MRST \cite{MRST2004}
sets of PDFs are based on the VFN schemes and contain a charm density. At the same time, the key
question remains open: How to measure the charm content of the proton? The basic theoretical
problem is that radiative corrections to the heavy-flavor production cross sections are large: 
they increase the leading order (LO) results by approximately a factor of two. Moreover, 
soft-gluon resummation of the threshold Sudakov logarithms indicates
that higher-order contributions can also be substantial. (For reviews, see 
Refs.~[\refcite{Laenen-Moch,kid1}].) On the other hand, perturbative instability leads to a high 
sensitivity of the theoretical calculations to standard uncertainties in the input QCD parameters: 
the heavy-quark mass, $m$, the factorization and renormalization scales, $\mu _{F}$ and $\mu _{R}$, 
$\Lambda_{\mathrm{QCD}}$ and the PDFs. For this reason, one can
only estimate the order of magnitude of the pQCD predictions for charm production cross sections in 
the entire energy range from the fixed-target experiments \cite{Mangano-N-R} to the RHIC collider 
\cite{R-Vogt}.

Since production cross sections are not perturbatively stable, they cannot be a good probe of the
charm density in the proton. For this reason, it is of special interest 
to study those observables that are well-defined in pQCD. Nontrivial examples of such observables were 
proposed in Refs.~[\refcite{we1,we2,we3,we4,we5,we7}], where the azimuthal $\cos(2\varphi)$ asymmetry and 
Callan-Gross ratio $R(x,Q^2)=F_L/F_T$  in heavy quark leptoproduction were analyzed.\footnote{Note also 
the recent paper [\refcite{Almeida-S-V}], where the perturbative stability of the QCD predictions 
for the charge asymmetry in top-quark hadroproduction has been observed.} 
It was shown that, contrary to the production cross sections, the azimuthal asymmetry \cite{we2,we4} and 
Callan-Gross ratio \cite{we7} in heavy flavor leptoproduction are stable within the FFNS, 
both parametrically and perturbatively.

\begin{figure}
\begin{center}
\mbox{\epsfig{file=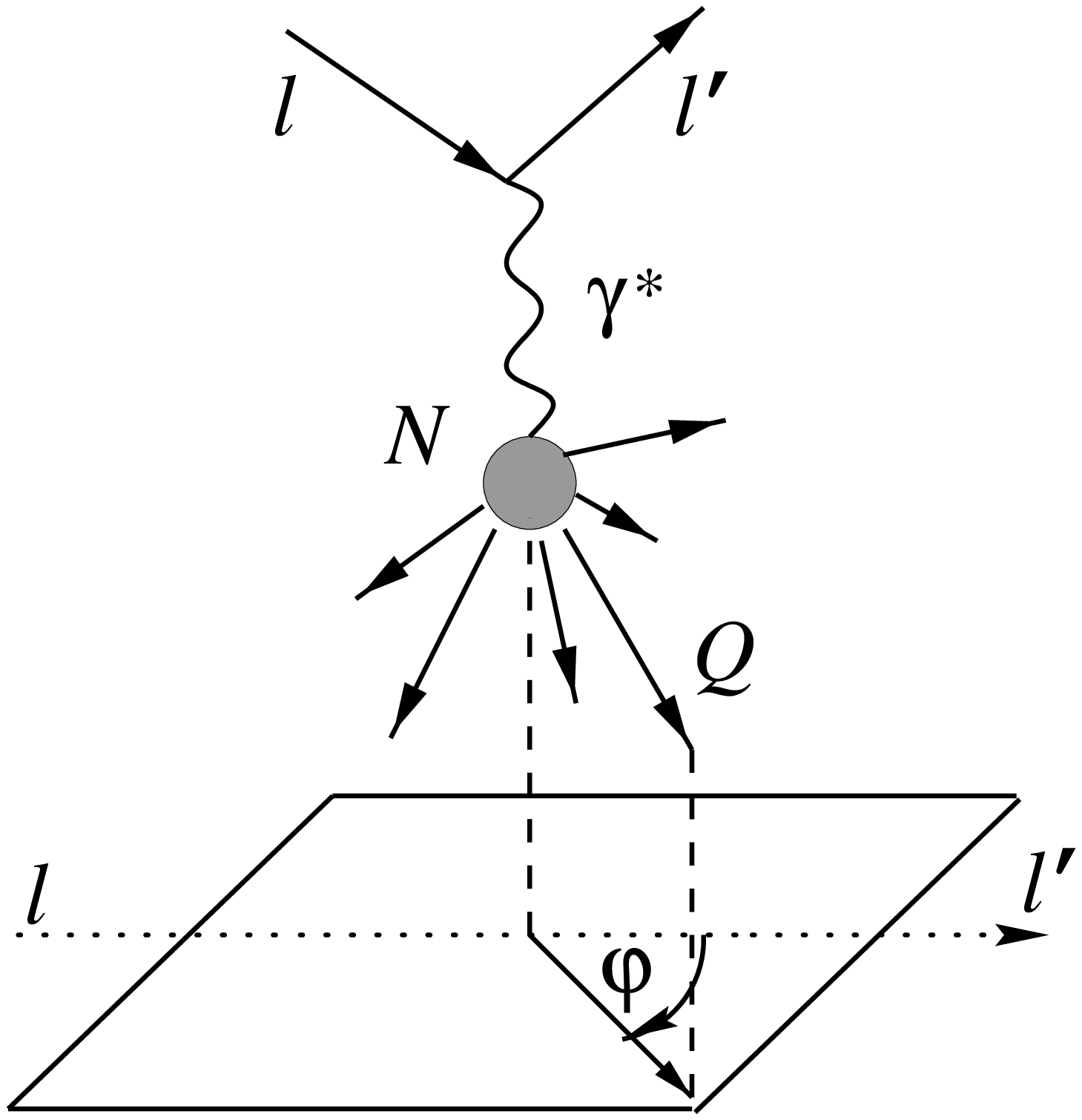,width=150pt}}
\caption{\label{Fig1}\small Definition of the azimuthal
angle $\varphi$ in the nucleon rest frame.}
\end{center}
\end{figure}

In the present talk, we discuss resummation of the mass logarithms of the type $\alpha_{s}\ln\left( Q^{2}/m^{2}\right)$ 
in heavy quark leptoproduction \cite{we5,we8}:
\begin{equation}
l(\ell )+N(p)\rightarrow l(\ell -q)+Q(p_{Q})+X[\overline{Q}](p_{X}). \label{1}
\end{equation}
The cross section of the reaction (\ref{1}) may be written as
\begin{eqnarray}
\frac{\mathrm{d}^{3}\sigma_{lN}}{\mathrm{d}x\mathrm{d}Q^{2}\mathrm{d}\varphi }&=&\frac{2\alpha^{2}_{em}}{Q^4}
\frac{y^2}{1-\varepsilon}\Bigl[ F_{T}( x,Q^{2})+ \varepsilon F_{L}(x,Q^{2}) \Bigr. \nonumber   \\
&+&\Bigl. \varepsilon F_{A}( x,Q^{2})\cos 2\varphi+
2\sqrt{\varepsilon(1+\varepsilon)} F_{I}( x,Q^{2})\cos \varphi\Bigr], \label{2}
\end{eqnarray}
where $F_{2}(x,Q^2)=2x(F_{T}+F_{L})$ while the quantity $\varepsilon$ measures the degree of the longitudinal 
polarization of the virtual photon in the Breit frame \cite{dombey}: $\varepsilon=\frac{2(1-y)}{1+(1-y)^2}$. 
The quantities $x$, $y$, and $Q^2$ are the usual Bjorken kinematic variables while the azimuth $\varphi$ is defined 
in Fig.~\ref{Fig1}.

In next sections, we will consider resummation of the mass logarithms for the quantities $R(x,Q^2)$ and $A(x,Q^2)$ 
defined as
\begin{equation}\label{3}
R(x,Q^{2})=\frac{F_{L}}{F_{T}}(x,Q^{2}), \qquad A(x,Q^{2})=2x\frac{F_{A}}{F_{2}}(x,Q^{2}).
\end{equation}

\section{\label{ratio} Resummation for $F_2$ and Callan-Gross Ratio}

\begin{figure}
\begin{tabular}{cc}
\mbox{\epsfig{file=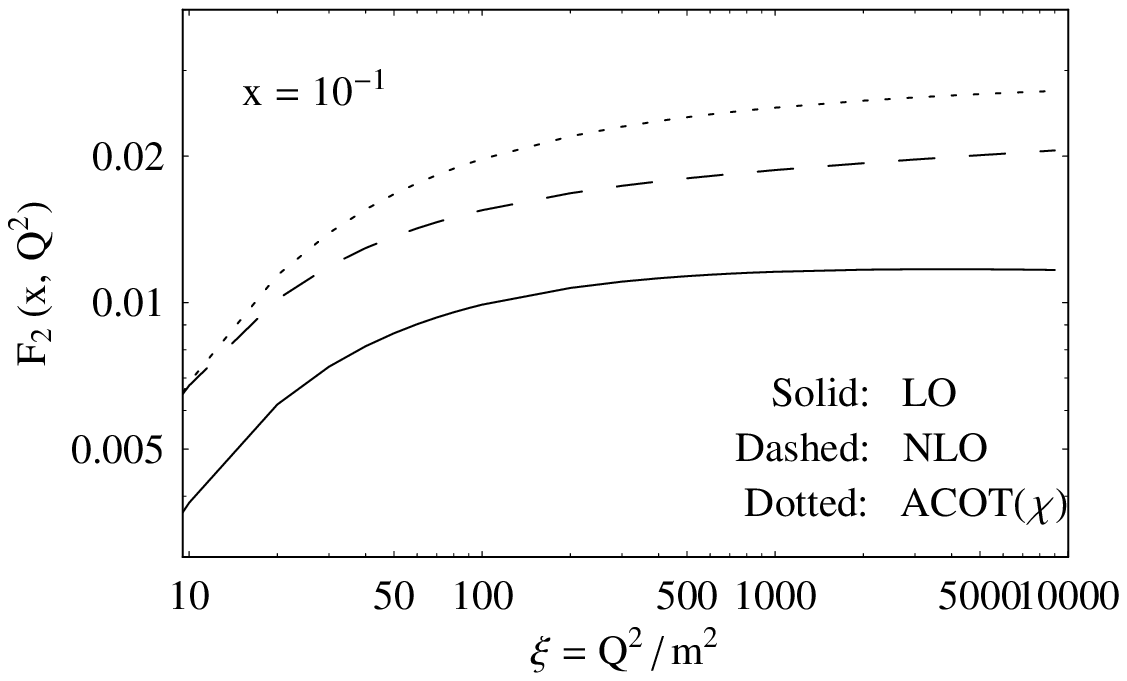,width=160pt}}
& \mbox{\epsfig{file=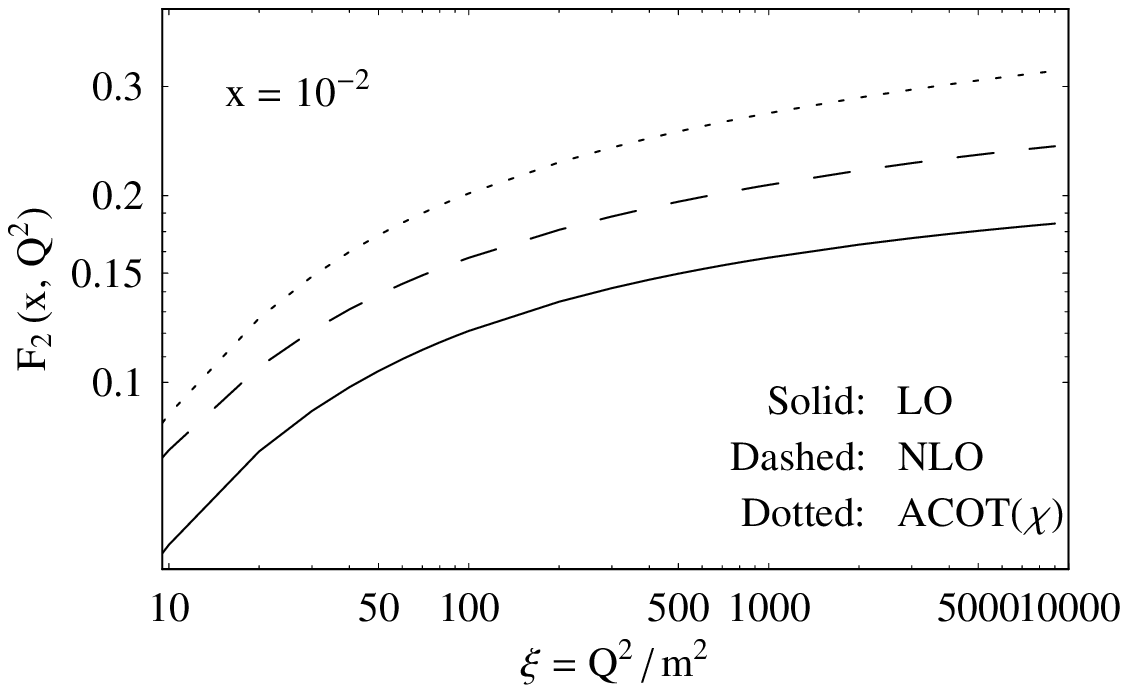,width=160pt}}\\
\mbox{\epsfig{file=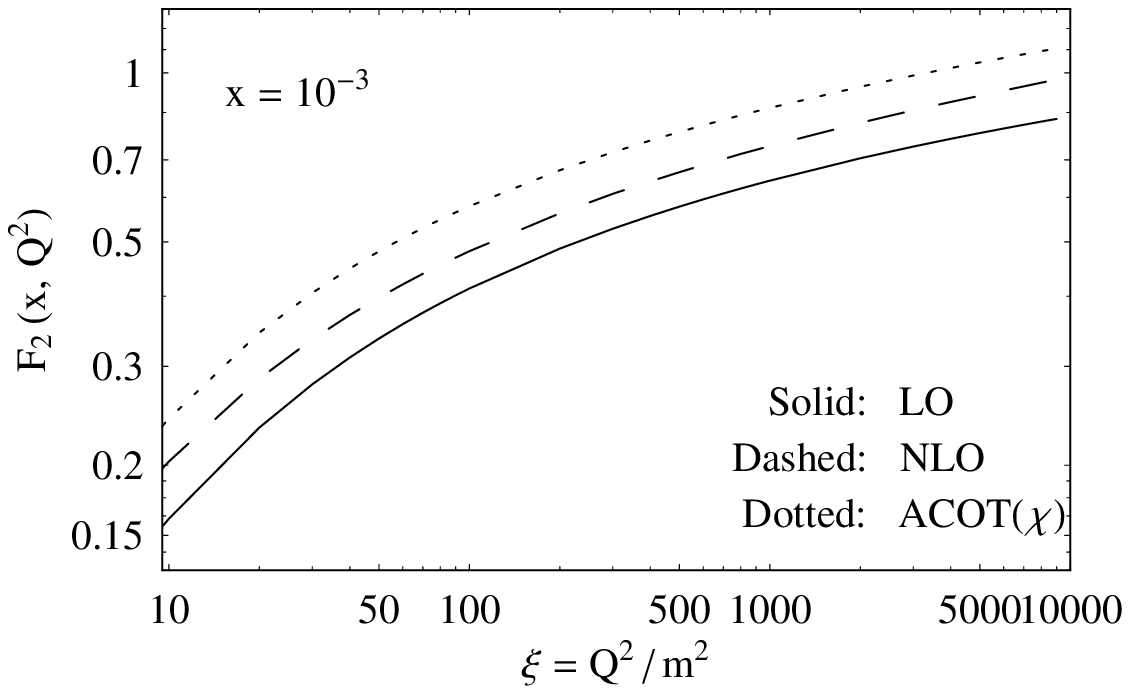,width=160pt}}
& \mbox{\epsfig{file=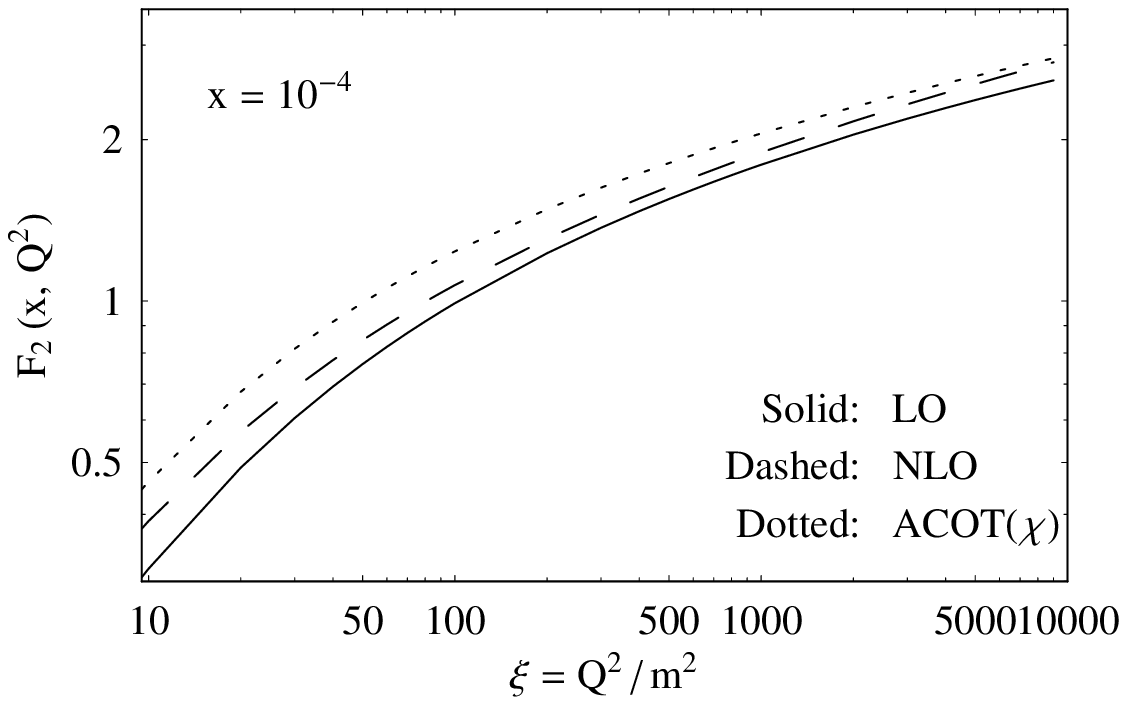,width=160pt}}\\
\end{tabular}
\caption{\label{Fig2}\small $Q^2$ dependence of the structure function $F_2(x,Q^2)$ in charm 
leptoproduction at $x=10^{-1}$, $10^{-2}$, $10^{-3}$ and $10^{-4}$. 
Plotted are the LO (solid lines) and NLO (dashed lines) FFNS predictions, as well as the 
ACOT($\chi$) VFNS (dotted curves) results.}
\end{figure}

\begin{figure}
\begin{tabular}{cc}
\mbox{\epsfig{file=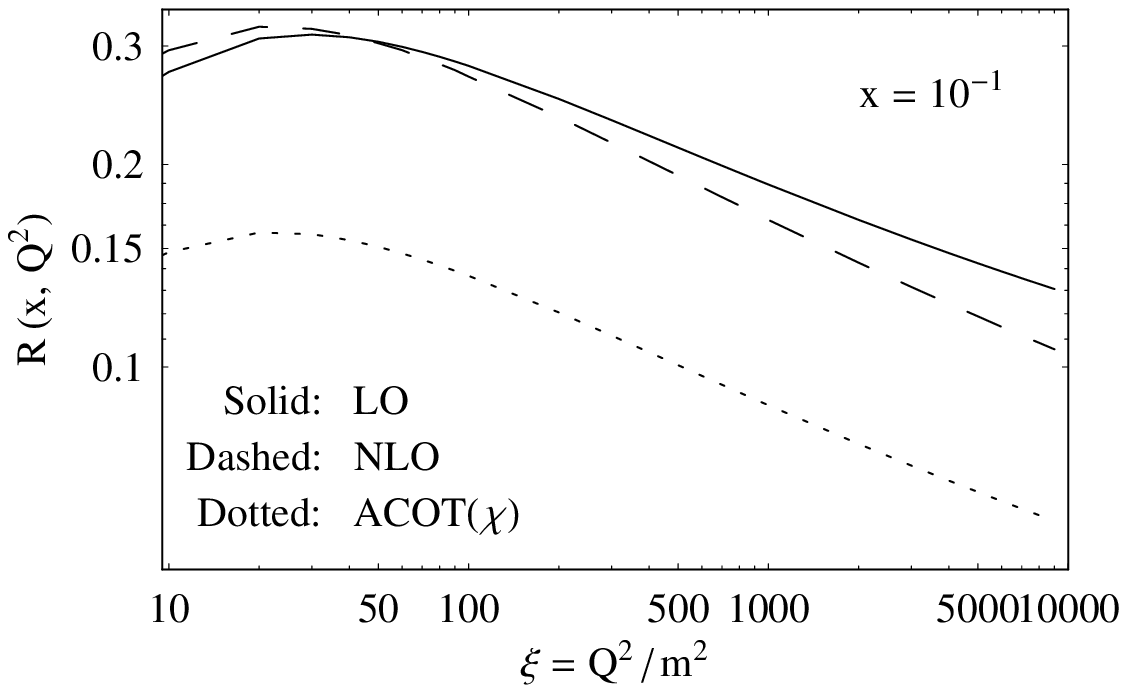,width=160pt}}
& \mbox{\epsfig{file=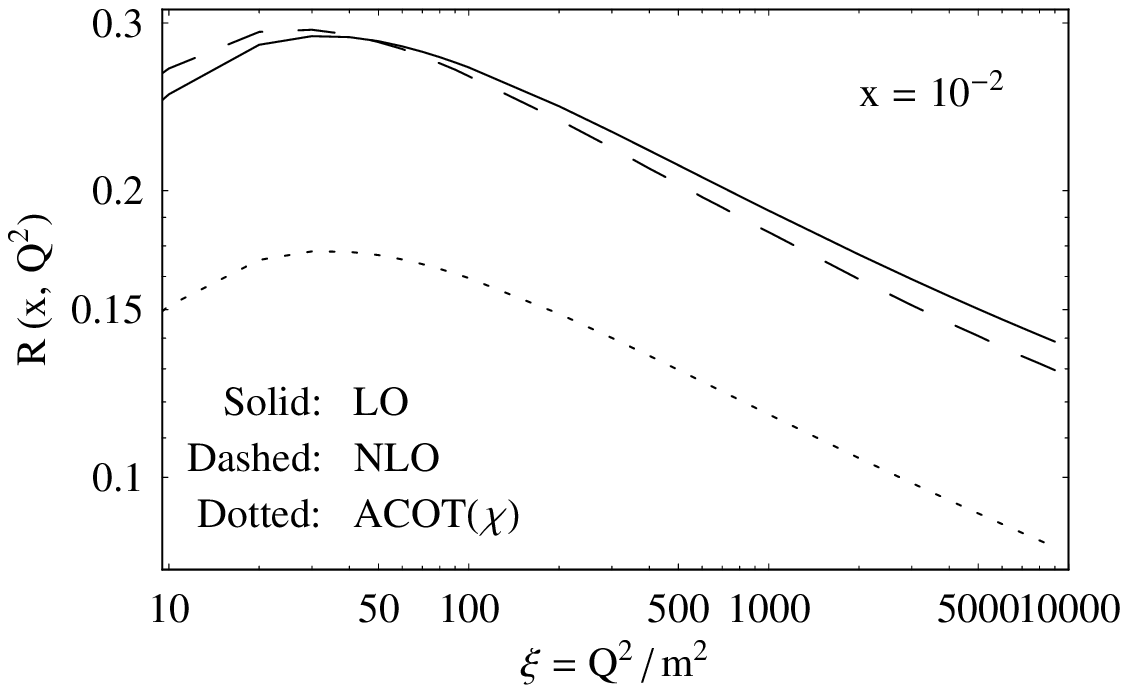,width=160pt}}\\
\mbox{\epsfig{file=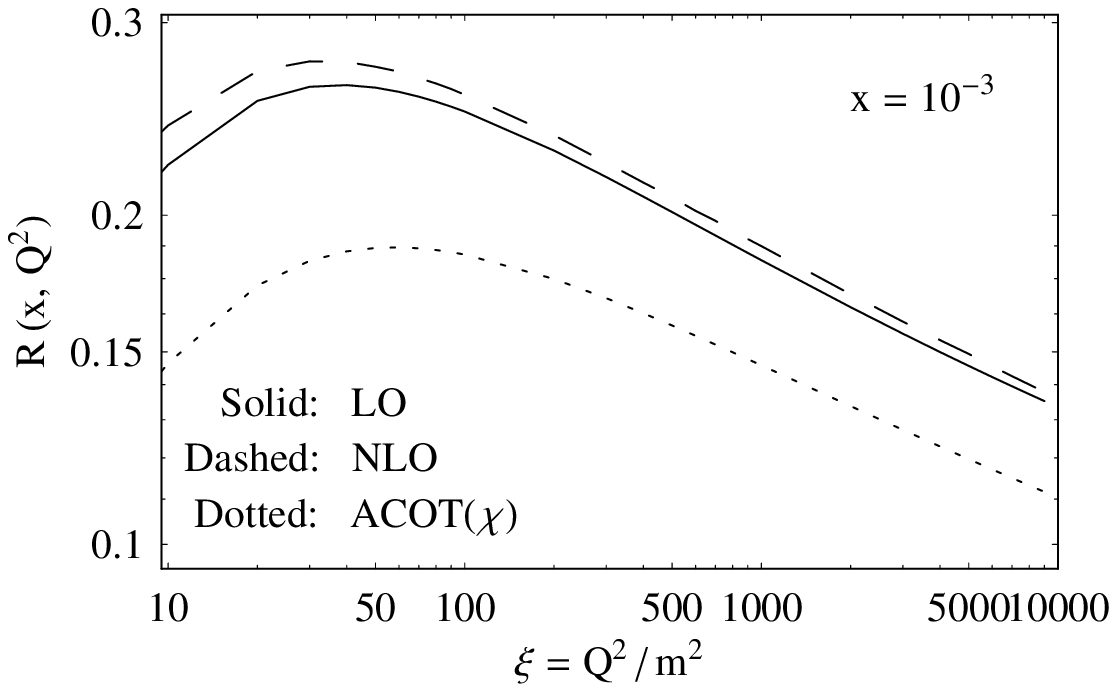,width=160pt}}
& \mbox{\epsfig{file=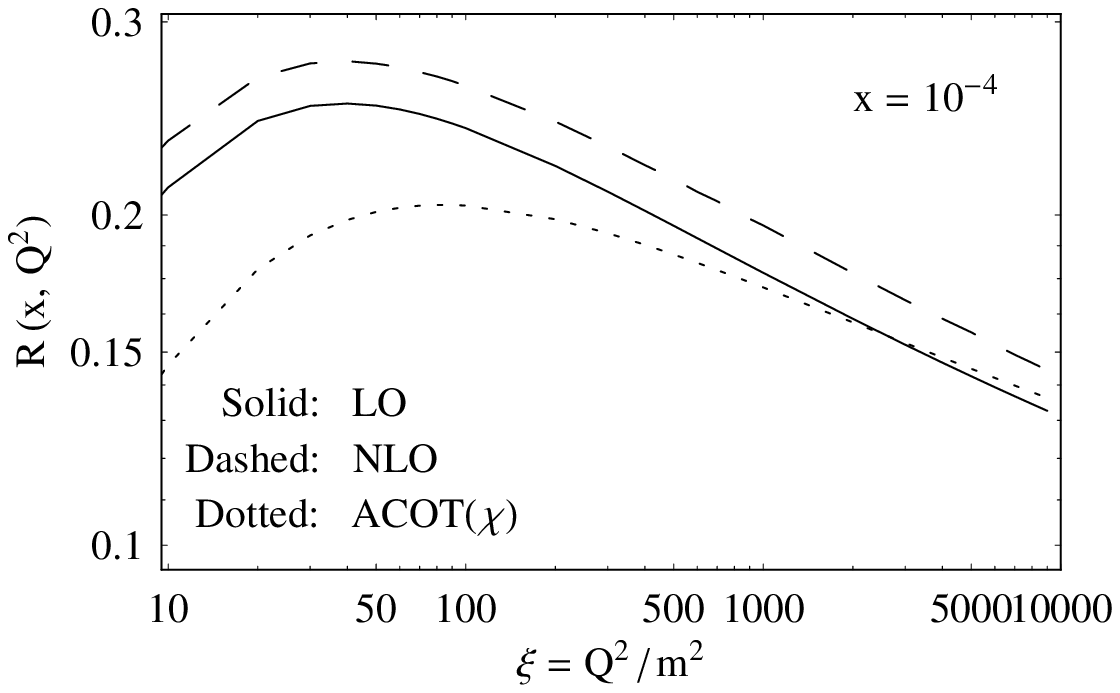,width=160pt}}\\
\end{tabular}
\caption{\label{Fig3}\small $Q^2$ dependence of the Callan-Gross ratio, $R(x,Q^2)=F_L/F_T$, in 
charm leptoproduction at $x=10^{-1}$, $10^{-2}$, $10^{-3}$ and $10^{-4}$. 
Plotted are the LO (solid lines) and NLO (dashed lines) FFNS predictions, as well as the 
ACOT($\chi$) VFNS (dotted curves) results.}
\end{figure}

To estimate the charm-initiated contributions, we use the ACOT($\chi$) VFNS proposed in 
Ref.~[\refcite{chi}].\footnote{For more details, see  Refs.~[\refcite{we5,we8}].} 
In Figs.~\ref{Fig2} and \ref{Fig3}, we present numerical analysis of the NLO corrections \cite{Bluemlein} and 
charm-initiated 
contributions to the structure function $F_{2}(x,Q^{2})$ and Callan-Gross ratio $R(x,Q^2)=F_L/F_T$ 
in charm leptoproduction. In our calculations, we use the CTEQ6M parametrization of the gluon 
and charm PDFs together with the value $m_c=1.3$~GeV [\refcite{CTEQ6}].\footnote{Note that we convolve 
the NLO CTEQ6M distribution functions with both the LO and NLO partonic cross sections 
that makes it possible to estimate directly the degree of stability of the FFNS predictions under 
radiative corrections.}
The default value of the factorization and renormalization scales is $\mu=\sqrt{4m_{c}^{2}+Q^{2}}$.

One can see from Fig.~\ref{Fig2} that, at $x\sim 10^{-1}$, both the radiative corrections and 
charm-initiated contributions to $F_{2}(x,Q^{2})$ are large: they increase the LO FFNS results by 
approximately a factor of two for all $Q^2$. At the same time, the relative difference between 
the dashed and dotted lines does not exceed $25\%$ for $\xi=Q^2/m^2<10^{3}$.

Considering the corresponding predictions for the ratio $R(x,Q^2)$ presented in Fig.~\ref{Fig3}, 
we see that, in this case, the NLO and charm-initiated contributions are strongly different. The NLO corrections
to $R(x,Q^2)$ are small, less than $15\%$, for $x\sim 10^{-3}$--$10^{-1}$ and $\xi<10^{4}$. On the other 
hand, the corresponding charm-initiated contributions are large: they decrease the LO FFNS predictions by 
about $50\%$ practically for all values of $\xi>10$. 
This is due to the fact that resummation of the mass logarithms has different effects on
the structure functions $F_{T}(x,Q^{2})$ and $F_{L}(x,Q^{2})$ because they have different 
dependences on the  quantities $\alpha_s^n\ln^k (Q^{2}/m^{2})$. In particular, contrary to 
the transverse structure function, $F_{T}(x,Q^{2})$, the longitudinal one, $F_{L}(x,Q^{2})$, 
does not contain potentially large mass logarithms at both LO and NLO \cite{LRSN,BMSMN}.
We conclude that, contrary to the the production 
cross sections, the Callan-Gross ratio $R(x,Q^2)=F_L/F_T$ could be good probe of the charm density in 
the proton at $x\sim 10^{-3}$--$10^{-1}$.

Note that this observation depends weakly on the PDFs we use. We have verified that all the recent CTEQ
versions \cite{CTEQ4,CTEQ5,CTEQ6} of the PDFs lead to a sizeable reduction of the LO FFNS 
predictions for the ratio $R(x,Q^2)$.

As to the low $x\to 0$ behavior of the Callan-Gross ratio, this problem requires resummation 
of the BFKL \cite{BFKL1} terms of the type $\ln (1/x)$  and will be considered in a forthcoming publication.

\section{\label{asymmetry} Resummation for Azimuthal Asymmetry}

\begin{figure}
\begin{tabular}{cc}
\mbox{\epsfig{file=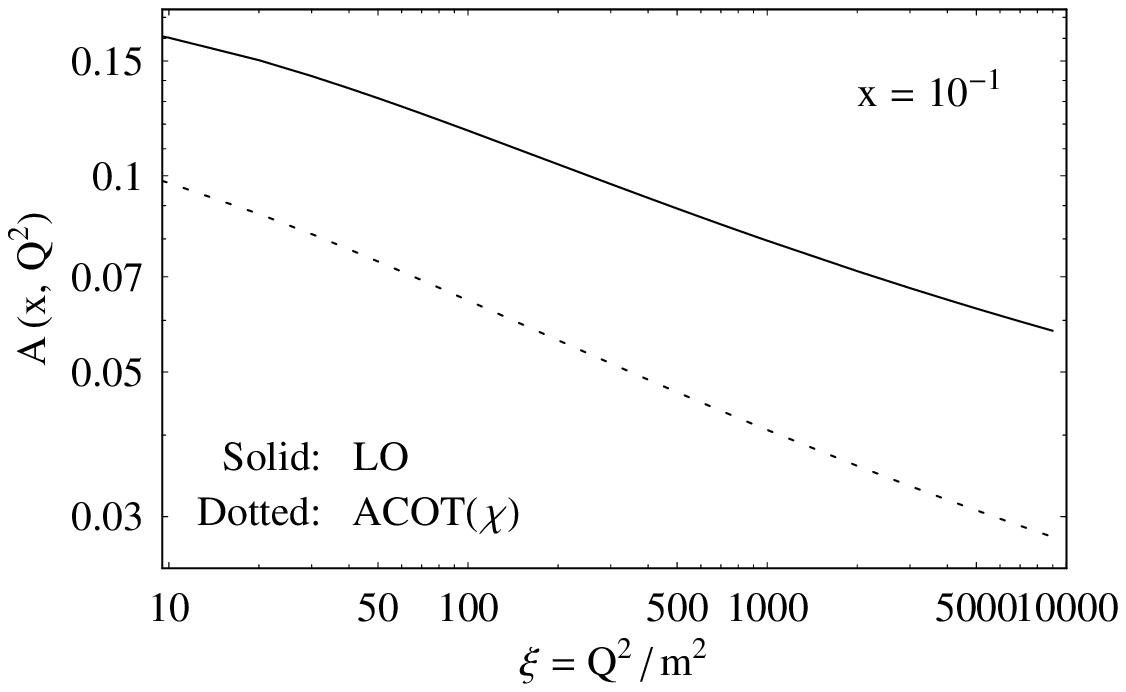,width=160pt}}
& \mbox{\epsfig{file=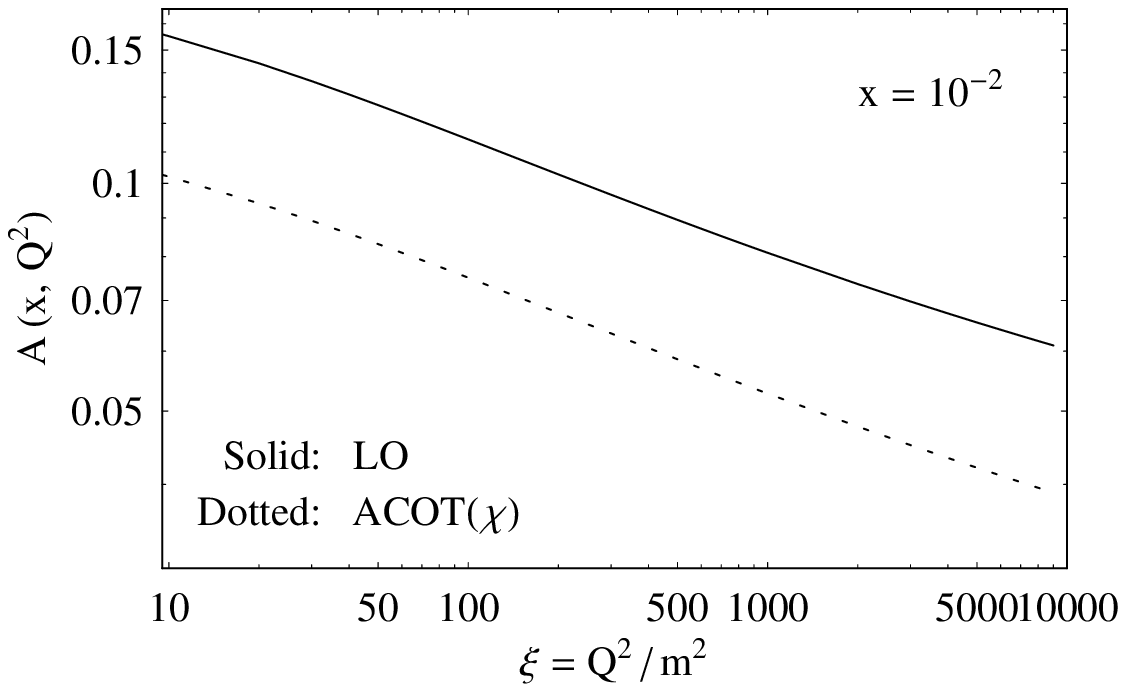,width=160pt}}\\
\mbox{\epsfig{file=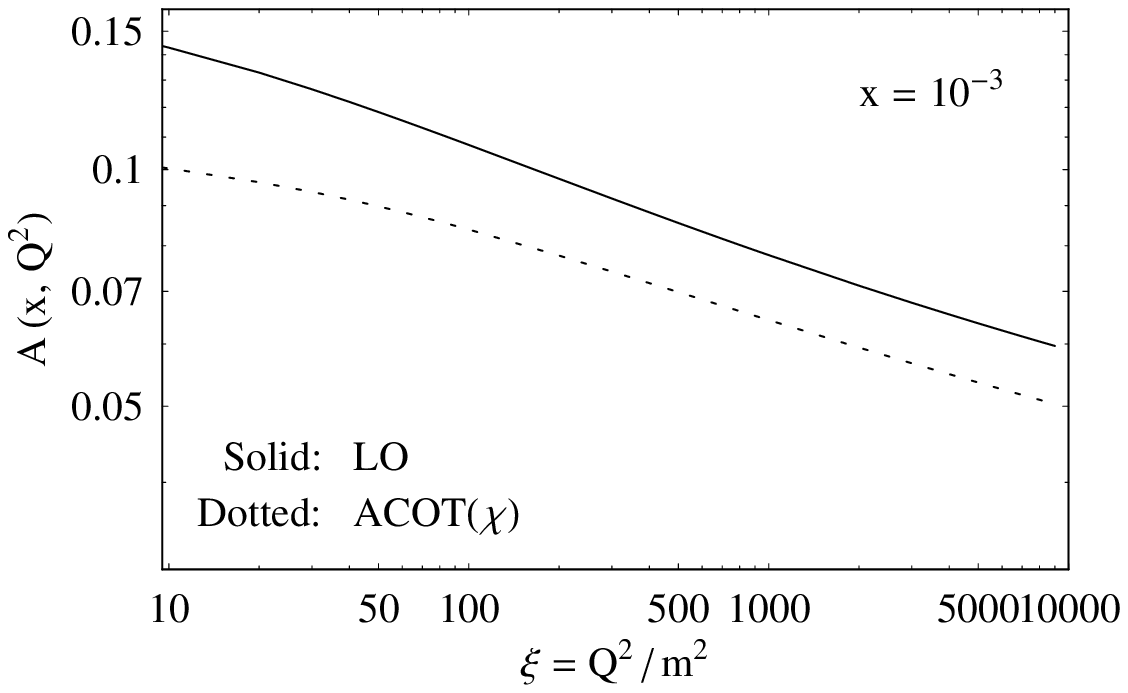,width=160pt}}
& \mbox{\epsfig{file=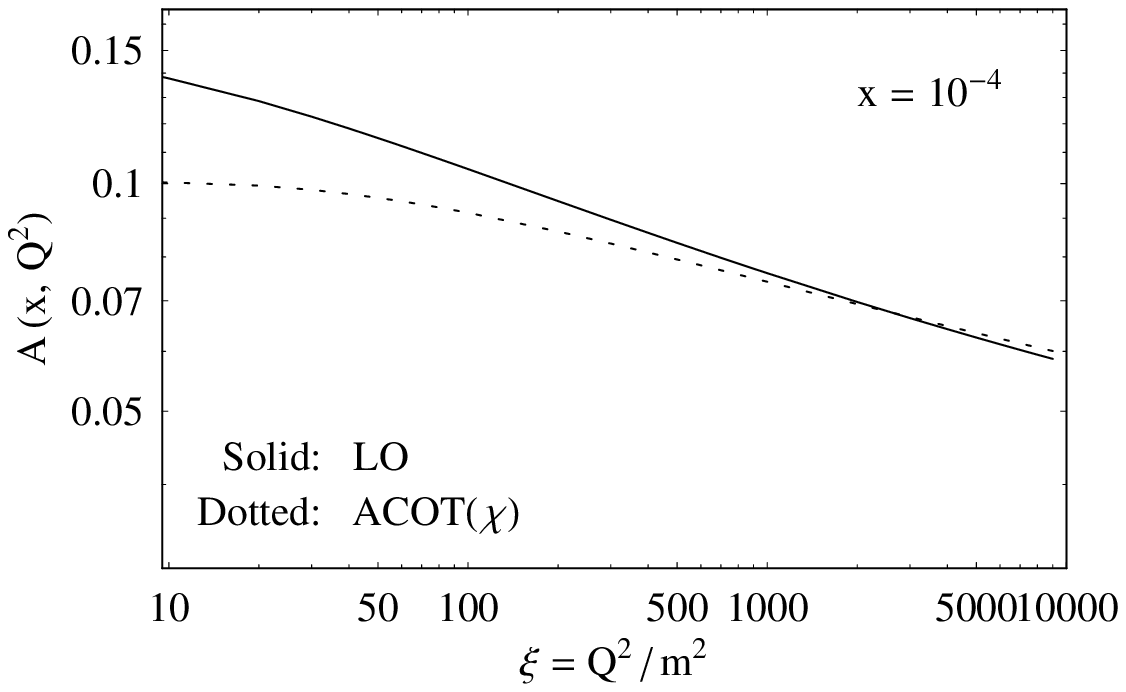,width=160pt}}\\
\end{tabular}
\caption{\label{Fig4}\small $Q^2$ dependence of the azimuthal asymmetry, $A(x,Q^{2})=2xF_{A}/F_{2}$, in 
charm leptoproduction at $x=10^{-1}$, $10^{-2}$, $10^{-3}$ and $10^{-4}$. 
Plotted are the LO FFNS (solid lines) and ACOT($\chi$) VFNS (dotted curves) results.}
\end{figure}

Fig.~\ref{Fig4} shows the ACOT($\chi$) predictions for the asymmetry parameter $A(x,Q^2)=2xF_{A}/F_{2}$ at
several values of variable $x$: $x=10^{-1}$, $10^{-2}$, $10^{-3}$ and $10^{-4}$. For comparison, we plot
also the LO FFNS predictions (solid curves). Again, we use the CTEQ6M parametrization of PDFs, $m_c=1.3$~GeV, 
and $\mu=\sqrt{4m_{c}^{2}+Q^{2}}$.

One can see from Fig.~\ref{Fig4} the following properties of the azimuthal asymmetry. 
The mass logarithms 
resummation leads to a sizeable decreasing of the LO FFNS predictions for the $\cos2\varphi$-asymmetry.
In the ACOT($\chi$) scheme, the charm-initiated contribution reduces the FFNS results for $A(x,Q^{2})$ by about
$(30$--$40)\%$ at $x\sim 10^{-2}$--$10^{-1}$. The origin of this reduction is the same as in the case of  $R(x,Q^2)$: 
contrary to $F_{2}$, the azimuth dependent structure function $F_{A}$ is safe in the limit $m^2\to 0$ at least at LO.

Presently, the exact NLO predictions for the azimuth dependent structure function $F_{A}$ are not available. However, 
in Ref.~[\refcite{we5}] the NLO corrections to the $\cos2\varphi$-asymmetry have been estimated within the so-called 
soft-gluon approximation at $Q^2 \lesssim m^2$.\footnote{The soft-gluon approximation is unreliable 
for high $Q^2\gg m^2$.} 
It was demonstrated that large soft-gluon corrections 
to both $F_{A}$ and $F_{2}$ cancel each other in their ratio $A=2xF_{A}/F_{2}$ with a good accuracy. 
For this reason, one can exepct that the $\cos2\varphi$-asymmetry is also stable, both parametrically and
perturbatively, in a wide kinematic range of variables $x$ and $Q^{2}$ within the FFNS.

We have also analyzed how the VFNS predictions depend on the choice of subtraction prescription. In
particular, the schemes proposed in Refs.~[\refcite{KS,SACOT}] have been considered. We have found that,
sufficiently above the production threshold, these subtraction prescriptions also reduce the LO FFNS results
for the asymmetry by approximately $(30$--$50)\%$.

One can conclude that impact of the  mass logarithms resummation on the $\cos2\varphi$ asymmetry is 
essential at $x\sim 10^{-2}$--$10^{-1}$ and therefore can be tested experimentally.

\section{\label{conclusion} Conclusion}

In the present talk, we compare the structure function $F_{2}$, Callan-Gross ratio $R=F_L/F_T$ and 
azimuthal asymmetry $A=2xF_{A}/F_{2}$ in charm leptoproduction as probes of the charm content of the proton. 
To estimate the charm-initiated contributions, we used the ACOT($\chi$) VFNS \cite{chi} and recent CTEQ sets \cite{CTEQ4,CTEQ5,CTEQ6} of PDFs. 
Our analysis of the radiative and charm-initiated corrections 
indicates that, in a wide kinematic range, both contributions to the structure function 
$F_{2}(x,Q^{2})$ have similar $x$ and $Q^2$ behaviors. For this reason, it is difficult 
to estimate the charm content of the proton using only data on $F_{2}(x,Q^{2})$.

The situation with the Callan-Gross ratio and azimuthal asymmetry seems to be more optimistic. 
Our analysis shows that resummation of the mass logarithms leads to reduction of the FFNS predictions for 
$R(x,Q^2)$ and $A(x,Q^2)$ by $(30$--$50)\%$ at $x\sim 10^{-2}$--$10^{-1}$ and $Q^2\gg m^2$. 
Taking into account the perturbative stability of the Callan-Gross ratio and 
azimuthal asymmetry within the FFNS \cite{we5,we8}, this fact implies that the 
charm density in the proton can, in principle, be determined from high-$Q^2$ data on 
$R=F_L/F_T$ and $A=2xF_{A}/F_{2}$.

Concerning the experimental aspects, the quantities $R(x,Q^{2})$ and $A(x,Q^{2})$ in charm leptoproduction 
can be measured in future studies at the proposed EIC \cite{eRHIC} and 
LHeC \cite{LHeC} colliders at BNL/JLab and CERN, correspondingly.

\section{Acknowledgments}

We thank S. I. Alekhin and J. Bl$\ddot{\rm u}$mlein for providing us with fast code \cite{Bluemlein} for numerical calculations 
of the NLO DIS cross sections. The author is grateful to H. Avakian, S.J. Brodsky and C. Weiss for useful discussions. 
This work was supported in part by the ANSEF 2010 grant PS-2033.

\end{document}